\documentclass[useAMS,usenatbib]{mn2e}
\usepackage{graphics}

\newcommand{\CtHt}{\mbox{C$_2$H$_2$}}

\def\oversim#1#2{\lower0.5pt\vbox{\baselineskip0pt \lineskip-0.5pt
     \ialign{$\mathsurround0pt #1\hfil##\hfil$\crcr#2\crcr\sim\crcr}}}

\begin{document}
  \title[Acetylene bands in LMC AGB stars]
  {Spitzer observations of acetylene bands in carbon-rich AGB stars 
  in the Large Magellanic Cloud}

\author[M. Matsuura et al.]
{M.~Matsuura$^{1,2,3}$, P.R.~Wood$^{4}$,  G.C.~Sloan$^{5}$,
A.A.~Zijlstra$^{1}$, J.Th.~van~Loon$^{6}$, 
\newauthor
M.A.T.~Groenewegen$^{7}$, 
J.A.D.L.~Blommaert$^{7}$, 
M.-R.L.~Cioni$^{8}$, 
M.W.~Feast$^{9}$, 
\newauthor
H.J.~Habing$^{10}$, 
S.~Hony$^{7}$, E. Lagadec$^{1}$, C.~Loup$^{11}$, J.W.~Menzies$^{12}$, 
\newauthor
L.B.F.M.~Waters$^{13,7}$,
P.A.~Whitelock$^{9,12,14}$\\
$^{1}$ School of Physics and Astronomy, University of Manchester, 
        Sackville Street, P.O. Box 88, Manchester M60 1QD, United Kingdom \\
$^{2}$ APS Division, Department of Pure and Applied Physics, 
        Queen's University Belfast, University Road, BT7 1NN, United Kingdom \\
$^{3}$ National Astronomical Observatory of Japan, Osawa 2-21-1, 
       Mitaka, Tokyo 181-8588, Japan \\
$^{4}$  Research School of Astronomy \& Astrophysics, Mount Stromlo Observatory,
        Australian National University, 
        Cotter Road, \\
        Weston ACT 2611, Australia \\
$^{5}$ Astronomy Department, Cornell University, 610 Space Sciences Building, 
        Ithaca, NY 14853-6801, USA \\
$^{6}$  Astrophysics Group, School of Physical and Geographical Sciences, Keele 
        University, Staffordshire ST5 5BG, United Kingdom \\
$^{7}$ Instituut voor Sterrenkunde, KU Leuven, Celestijnenlaan 200B, 
        3001 Leuven, Belgium \\
$^{8}$  SUPA, School of Physics, University of Edinburgh, IfA, 
        Blackford Hill, Edinburgh EH9 3HJ, United Kingdom \\
$^{9}$  Astronomy Department, University of Cape Town, 7701 Rondebosch, 
        South Africa \\
$^{10}$  Sterrewacht Leiden, Niels Bohrweg 2, 2333 RA Leiden, The Netherlands \\
$^{11}$ Institut d'Astrophysique de Paris, CNRS, 98bis Boulevard Arago, 75014 Paris, France \\
$^{12}$ South African Astronomical Observatory, P.O.Box 9, 7935     
        Observatory, South Africa \\
$^{13}$ Astronomical Institute ``Anton Pannekoek'', University of Amsterdam, 
        Kruislaan 403, 1098 SJ, Amsterdam, \\
        The Netherlands \\
$^{14}$ NASSP, Department of Mathematics and Applied Mathematics, 
        University of Cape Town, 7701 Rondebosch, South Africa \\
             }

\date{Accepted. Received; in original form }
\pagerange{\pageref{firstpage}--\pageref{lastpage}} \pubyear{2006}

\maketitle
\label{firstpage}

\begin{abstract}
  We investigate the molecular bands in carbon-rich AGB stars in the Large
  Magellanic Cloud (LMC), using the InfraRed Spectrograph (IRS) on board the
  {\it Spitzer Space Telescope} (SST) over the 5--38~$\mu$m range.  All 26
  low-resolution spectra show acetylene (\CtHt) bands at 7 and
  14~$\mu$m.  The hydrogen cyanide (HCN) bands at these wavelengths are very
  weak or absent. This is consistent with low nitrogen abundances in the
  LMC.  The observed 14~$\mu$m \CtHt\, band is reasonably reproduced by an
  excitation temperature of 500~K. There is no clear dilution of the
  14~$\mu$m \CtHt\, band by circumstellar dust emission.  This 14~$\mu$m
  band originates from molecular gas in the circumstellar envelope in these
  high mass-loss rate stars, in agreement with previous findings for
  Galactic stars. The \CtHt\,column density, derived from the 13.7~$\mu$m
  band, shows a gas mass-loss rate in the range $3\times 10^{-6}$ to
  $5\times 10^{-5} M_{\sun}$\,yr$^{-1}$.  This is comparable with the total
  mass-loss rate of these stars estimated from the spectral energy
  distribution.  Additionally, we compare the line strengths of the
  13.7~$\mu$m \CtHt\, band of our LMC sample with those of a Galactic
  sample.  Despite the low metallicity of the LMC, there is no clear
  difference in the \CtHt\, abundance among LMC and Galactic stars. This
  reflects the effect of the 3rd dredge-up bringing self-produced carbon to
  the surface, leading to high C/O ratios at low metallicity.
\end{abstract}

%
\large


\section{ Introduction}
Low and intermediate mass stars (i.e. those with main sequences masses of
1--8\,$M_{\sun}$) experience intensive mass loss towards the end of their
life.  During the asymptotic giant branch (AGB) phase, these stars lose up
to 90\% of their initial mass. As their population is large, AGB stars are
among the most important sources of gas and dust in the local universe.

Material lost from the star forms a circumstellar envelope, which consists
of gas and dust grains.  In this circumstellar envelope, various kinds of
molecules exist. The molecules first form in the stellar photosphere.  In
the circumstellar envelope, an active chemistry, particularly in regions
where interstellar UV radiation penetrates, leads to further molecules
\citep{Millar00}.  However, the structure of the circumstellar envelope and
the formation region of these molecules are not yet completely understood.
Molecules are also formed in the innermost regions of the envelope where the
stellar pulsations set up shock waves \citep{Duari00, Gautschy-Loidl04}.
Around carbon-rich stars, which have a carbon to oxygen ratio of C/O$>$1,
CO, HCN and \CtHt\, and many carbon chain molecules are detected. \CtHt,
which is the focus of this paper, is one of the most abundant molecules
after CO in carbon-rich envelopes.  Note in particular that \CtHt\, may be
important for dust formation in carbon stars
\citep{Keady88}.

The majority of nearby galaxies have a lower metallicity than the Milky Way.
The metallicity of red giants in the Galactic disk ranges from
[Fe/H]=$-0.11$ to +0.06 \citep{Smith85}. The mean metallicity of the Large
Magellanic Cloud (LMC) is about half of the solar metallicity. The
metallicity for LMC AGB stars is uncertain, but may be similar to that found
in red giants from [Fe/H]=$-1.1$ to [Fe/H]=$-0.3$ \citep{Smith02}.  A lower
metallicity should affect the abundances of molecules in cool stars. However,
for AGB stars, carbon atoms are synthesised in the core and brought to the
surface by the third dredge-up. Therefore, abundances of carbon-bearing
molecules depend on the balance of the low metallicity effect and the
enrichment of carbon atoms through nuclear synthesis.  Previous studies
\citep{Matsuura02b, Matsuura05} using 3-$\mu$m spectra have found that the
\CtHt\, abundance is higher in carbon-rich stars in the LMC than in similar
stars our Galaxy.  This was first suggested by a study of the 3.1~$\mu$m
HCN+\CtHt\,band \citep{vanLoon99b}. However, 3-$\mu$m \CtHt\, is a mixture
of photospheric origin and circumstellar origin \citep{vanLoon06}, and the
influence of the temperature of the central star is not negligible.  In this
study, we use {\it Spitzer} mid-infrared spectra covering the 13.7\,$\mu$m
\CtHt\, bands. These bands are predominantly of circumstellar origin and,
therefore, we are able to study the molecular abundance in the circumstellar
envelope, including the gas mass-loss rate and metallicity effects.  The
{\it Spitzer Space Telescope} enables us to carry out a study of the
13.7\,$\mu$m \CtHt\, band in AGB stars in the LMC, for the first time.

\section{ Observations and Analysis}
 
Spectra of carbon stars in the LMC have been obtained using the InfraRed
Spectrograph \citep[][IRS]{Houck04} on board the {\it Spitzer Space
Telescope} (SST).  The stars were observed as part of the General Observing
time of SST, programme 3505 (P.I.  P.R.  Wood).  Two low-resolution modules
Short-Low (SL) and Long-Low (LL) were used, which provide wavelength
coverage from 5--38\,$\mu$m and spectral resolution from $R$=70--130. 
Spectra were covered by four segments, namely SL 2nd order
(5.3--8.7\,$\mu$m), SL 1st order (7.4--14.5\,$\mu$m), LL 2nd order
(14.0--21.3\,$\mu$m) and LL 1st order (19.5--38.0\,$\mu$m). In addition,
extra-spectra were obtained with `bonus' segments (7.3--8.8 and
20--22\,$\mu$m). The 7-$\mu$m C$_2$H$_2$ bands fall on two segments, which
could leave some uncertainty in the overlap region.
\citep{Zijlstra06} provide an overview of the project and describe the details
of target selection, observations and data-reduction.



For the analysis, we calculate synthetic spectra of HCN and \CtHt\,
molecular bands using the line lists from the HITRAN database, 2004 edition
\citep{Jacquemart03, Rothman05}.  The partition function of \CtHt\, is
available up to a temperature of 500\,K, therefore, the modelling is limited
to this temperature range.  There are insufficient lines in the model at
7~$\mu$m and this band cannot therefore be reproduced.  Calculation codes
are described in \citet{Matsuura02a}; a slab local thermodynamic equilibrium
(LTE) model with a single excitation temperature and a single column density
is used. The model is intended to identify the molecular bands and to
estimate their approximate excitation temperatures and density. Thus, this
procedure does not reveal fully the atmospheric structure of the AGB stars.
One of the uncertainties is any effect due to spherically symmetric
structure. In the slab model, molecular emission from the circumstellar
envelope is not considered, thus, the column densities tend to be higher
solved for this slab model than they would be for spherical symmetric
structure.

\section{ Description of spectra} 

\begin{figure}
\centering
\resizebox{\hsize}{!}{\includegraphics*{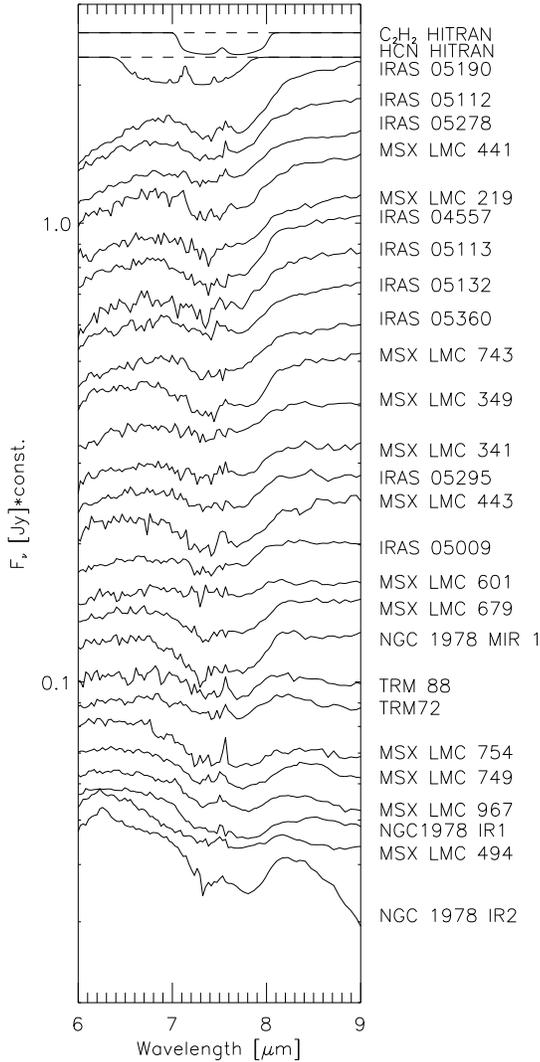}}
\caption{
Spitzer spectra around 7.5~$\mu$m of LMC carbon stars,
showing the \CtHt\, band. The spectra are sorted in order of the 
decreasing [6.4]$-$[9.3] infrared colour from top to bottom.
Model spectra of \CtHt\, and HCN are used to identify the 
molecular bands (\CtHt\, absorption is amplified in intensity by a factor of 2.5).
}
\label{Fig-spec7}
\end{figure}
\begin{figure}
\centering
\resizebox{\hsize}{!}{\includegraphics*{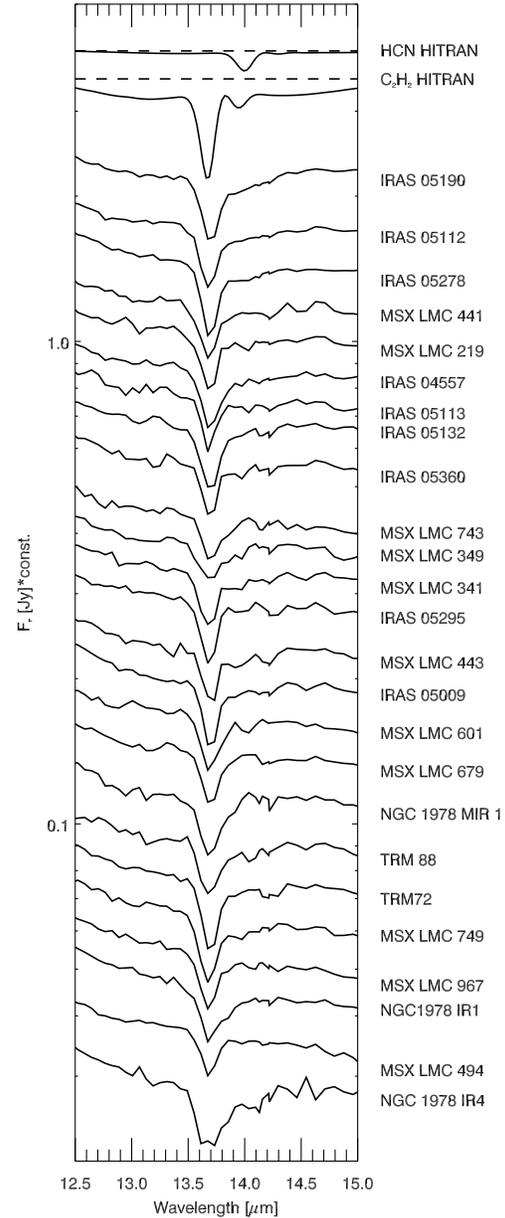}}
\caption{Acetylene band 
at 13.7~$\mu$m from the Spitzer spectra, in the same order 
as Fig.~\ref{Fig-spec7}.
}
\label{Fig-spec14}
\end{figure}

Fig.~\ref{Fig-spec7} shows the 7\,$\mu$m region of the Spitzer spectra,
arranged in order of the infrared colour [6.4]$-$[9.3]. From bottom to top,
the colour becomes redder (hereafter, `redder' means that the ratio of
long-wavelength to short-wavelength flux is larger).  The definition of the
colour (the `Manchester system') can be found in \citet{Zijlstra06} and
\citet{Sloan06}.  A double-peaked molecular absorption band centred at
7.5\,$\mu$m is found in all of the spectra. This absorption is associated
with the \CtHt\,
$\nu_4^1+\nu_5^1$ bands (P- and R-branches), which are found in the model
spectrum of \CtHt.
 A synthetic spectrum of \CtHt\, is plotted at the top of Fig.~\ref{Fig-spec7}.
(The model uses an excitation temperature of $T_{\rm
  ex}=$500~K, and a column density of $n=1\times 10^{20}$\,cm$^{-2}$.) 
There are insufficient \CtHt\, lines included in HITRAN to reproduce the shape 
(depth and width)
of this feature. Nevertheless, the synthetic spectrum confirms the
identification of this feature from the wavelength. Because of the
incompleteness of the line list at this wavelength range, the column density
derived from the 7~$\mu$m feature is not reliable.

Around 7\,$\mu$m, other molecular bands have been found in Galactic carbon
stars.  \citet{Aoki98} reported a SiS first overtone band at 6.6--7\,$\mu$m.
This band is relatively weak (less than 5 \% absorption with respect to the
continuum, except for a few contributing lines) according to \citet{Aoki98},
and such a weak band cannot be clearly detected in our LMC spectra.  The HCN
$2 \nu_2^{~0}$ P-branch and R-branch are found at about 6.5--7.7\,$\mu$m
\citep{Carter93, Rothman05}.  Fig.~\ref{Fig-spec7} also includes the HCN
model profile. This molecule may be responsible for the shallow suspected
absorption in NGC 1978 IR4 and MSX LMC 494.  Except for these cases, the
presence of HCN bands is not clear at low wavelength resolution.  CS
fundamental bands are found at 7--8\,$\mu$m with a bandhead at 7.3\,$\mu$m
\citep{Aoki98}. This feature is not detected in the spectra discussed here.
This is probably because CS is masked by strong \CtHt\, bands, while the
\citet{Aoki98} sample consists of relatively warm carbon stars and \CtHt\,
molecules are not formed efficiently in the hotter photospheres
\citep{Tsuji81}.

Fig.~\ref{Fig-spec14} shows the 14\,$\mu$m region of the Spitzer spectra
sorted in the same order as in Fig.~\ref{Fig-spec7}.  A sharp absorption
band at 13.7\,$\mu$m is found in all of the spectra.  This feature is due to
the Q-branch band of the \CtHt\,fundamental $\nu_5$ and other transitions
involving $\nu_5$.  The P- and R-branches of \CtHt\, produce broad
absorption features in the synthetic spectrum on either side of
the Q-branch band. There is a second peak of \CtHt\, absorption at
13.9~$\mu$m, and this band may also be found in the observed spectra.  As a
reference, we show the model spectra of \CtHt\,
($n=1\times10^{18}$\,cm$^{-2}$, $T_{\rm ex}=500$\,K) and HCN
($n=3\times10^{17}$\,cm$^{-2}$, $T_{\rm ex}=500$\,K).

\begin{figure}
\centering
\resizebox{\hsize}{!}{\includegraphics*{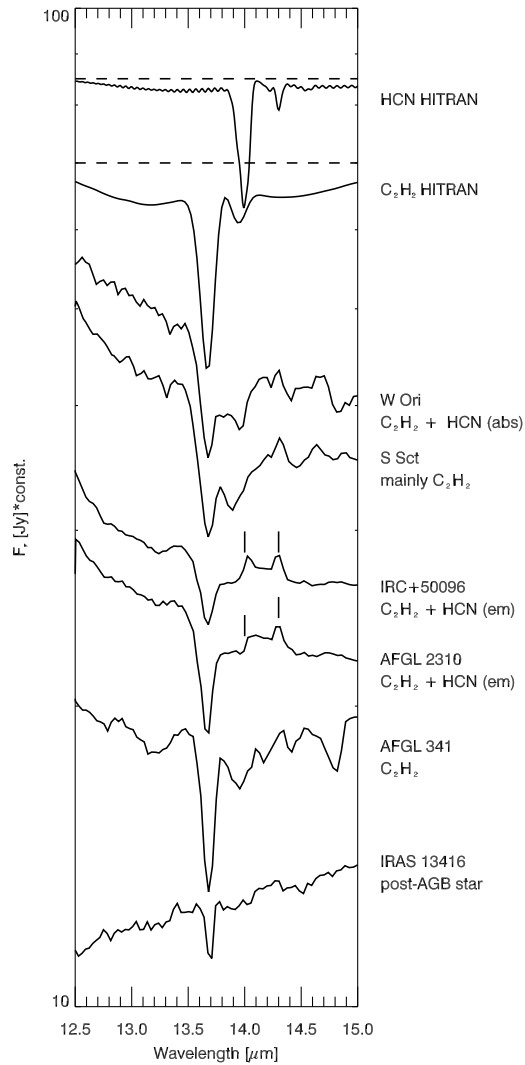}}
\caption{Representative spectra of the \CtHt\, and the HCN
bands found in the ISO spectra of Galactic carbon stars.
}
\label{Fig-spec14ISO}
\end{figure}

\citet{Cernicharo99} and \citet{Aoki99} studied ISO spectra of Galactic carbon
stars: they found that two molecular bands are blended in the 13--14\,$\mu$m
range, namely \CtHt\,$\nu_5$ (and other transitions
involving $\nu_5$) and HCN fundamental $\nu_2^{~1}$.
Fig.~\ref{Fig-spec14ISO} shows some example ISO/SWS spectra.
\citet{Hony02} initially presented these ISO/SWS data, 
illustrating AGB stars and planetary nebulae.  The ISO/SWS AOT1 resolution
of $R\sim300$ is higher than that of the Spitzer/IRS. The characteristic
absorption features of these two molecules are clearly separated.  The
\CtHt\, band is centred at 13.7\,$\mu$m, while the sharp absorption of HCN
is found at 14\,$\mu$m.  In addition, the HCN bands are occasionally found
in emission \citep{Aoki99, Cernicharo99}, as seen in IRC+50\,096 and
AFGL~2310 (Fig.~\ref{Fig-spec14ISO}).

From the ISO/SWS spectra, we define representative spectra, which show (1)
\CtHt\,only (2) \CtHt + HCN in absorption (3) \CtHt + HCN in emission. Within
category (1), there are two sub-categories, which contain hot
\CtHt\, traced by a  prominent 13.9\,$\mu$m feature \citep{Aoki99}, and
cold \CtHt\, without that feature.

 Among our LMC spectra, there is a possible detection of 14.0\,$\mu$m HCN
absorption in TRM~88 and MSX LMC 601. Apart from these sources, there is no
clear detection of HCN either in emission or in absorption.  In addition,
the 13.9\,$\mu$m \CtHt\, absorption is also weak; there is no LMC spectrum
with \CtHt\, comparable to the Galactic carbon star S~Sct, which shows a
strong secondary \CtHt\ absorption feature at 13.9\,$\mu$m in addition to
the broad feature at 13.7\,$\mu$m \CtHt.  This shows that \CtHt\, in our LMC
sample is relatively cold.

\begin{figure}
\centering
\resizebox{\hsize}{!}{\includegraphics*{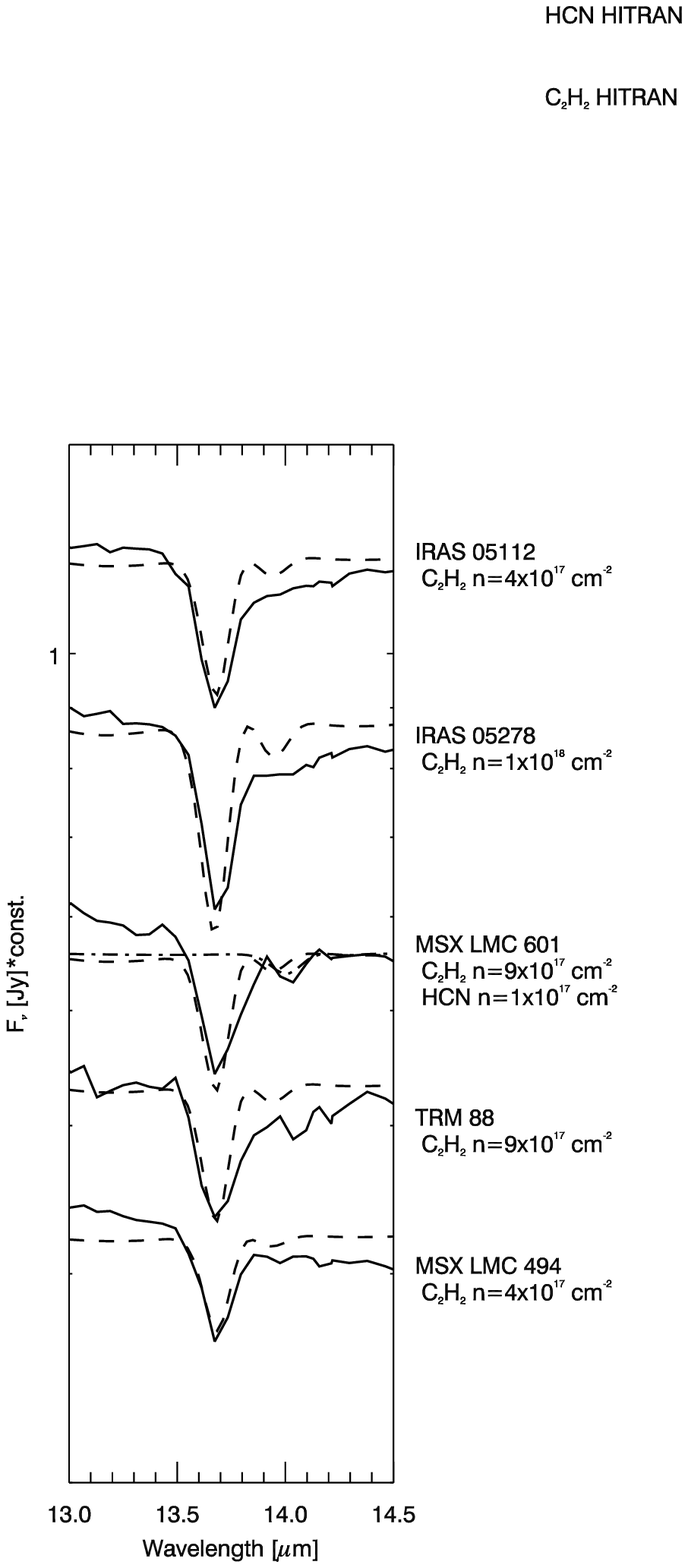}}
\caption{
The model fit to some of the spectra at 14~$\mu$m.
The solid lines are observed spectra, and dash lines are for 
the \CtHt\, model and dash-dot line is for HCN.
The excitation temperature of \CtHt\, is 500~K for all.
}
\label{Fig-c2h2-model}
\end{figure}

In Fig.~\ref{Fig-c2h2-model}, we fit representative LMC spectra with the
model. The column density of \CtHt\, is
$4\times10^{17}$--$1\times10^{18}$~cm$^{-2}$ (Table~\ref{table-model}). A
model spectrum of HCN is also added to \CtHt\, spectrum for a comparison
with MSX LMC 601 spectrum. The column density of HCN is
$1\times10^{17}$~cm$^{-2}$.


\begin{table}
\caption{Model parameters}
\begin{tabular}{lllll}
\hline
Name & Molecule & Column density \\
\hline
IRAS~05278$-$6942 & \CtHt & $1\times 10^{18}$ \\
IRAS 05112$-$6755 & \CtHt & $4\times 10^{17}$ \\
MSX~LMC~494       & \CtHt & $4\times 10^{17}$ \\
MSX~LMC~601       & \CtHt & $9\times 10^{17}$ \\
                  & HCN   & $1\times 10^{17}$ \\
TRM~88            & \CtHt & $9\times 10^{17}$ \\ \hline

\end{tabular}
\label{table-model}
\end{table}

\section{ Non-dilution}

 The spectra in Fig.~\ref{Fig-spec14} are sorted in order of colour,
[6.4]$-$[9.3], so that the colour increases from bottom to top.  Note
particularly that the 13.7\,$\mu$m Q-branch band does not fade away towards
redder stars.  
The [6.4]-[9.3] colour increases mainly as the contribution of the dust emission increases 
relative to the stellar emission.
The [6.4]$-$[9.3] colour is a measure of the optical depth
of the shell, and therefore of the mass-loss rate; the 13.7\,$\mu$m
\CtHt\, does not weaken towards more highly obscured stars, i.e., higher
mass-loss rate stars.

This result is in contrast to other \CtHt\, bands, such as the
3.1\,$\mu$m HCN+\CtHt\, band and the 3.8\,$\mu$m \CtHt\ band
\cite[e.g.][]{vanLoon99b,vanLoon06,
Matsuura02b,Matsuura05} where the 3-$\mu$m band equivalent width is observed
to be smaller in redder carbon stars. This difference between the 13-$\mu$m
and 3-$\mu$m bands cannot be a consequence of differences between the stars
examined.  Both this work and \citet{Matsuura05} discuss stars with similar
near-infrared colours ($1<H-K<3$, and $0.3<H-K<3.2$ respectively).
IRAS~05112$-$7655 was indeed observed here and in \citet{vanLoon99b} and
\citet{Matsuura05}. Therefore, the 13.7~$\mu$m \CtHt\, band behaves in a
different way with mass-loss rate from the bands at shorter wavelength.  The
decreasing equivalent width of the 3.8\,$\mu$m bands can be explained if
dust emission fills in the absorption band. There is no obvious evidence of
such dilution by dust within the 13.7\,$\mu$m \CtHt\, band, which suggests
that this band is formed throughout the circumstellar envelope.

\begin{figure}
\centering
\resizebox{\hsize}{!}{\includegraphics*{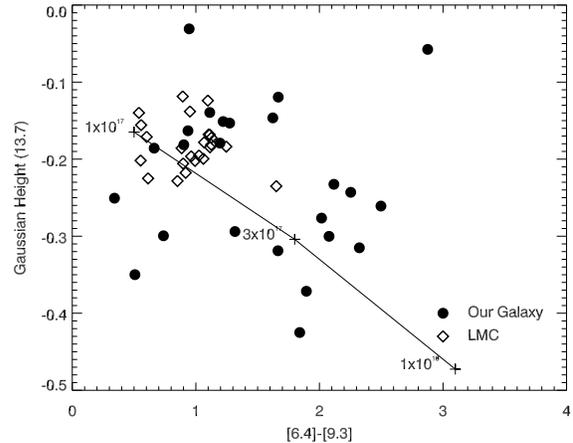}}
\caption{
The comparison of band strength (height of Gaussian fit at 13.7~$\mu$m) for
LMC stars (diamonds) and Galactic stars (filled circles). The Gaussian depth
is relative to the continuum. There is no clear difference in line strength
between Galactic and LMC samples. This band tends to become stronger towards
redder stars, i.e. higher mass-loss rate stars. The crosses show the \CtHt\,
model band strength at the excitation temperature of 500~K and the column
density of $1\times 10^{17}$, $3\times 10^{17}$, and
$1\times 10^{18}$~cm$^{-2}$. The horizontal scale of the crosses
is arbitrary and is given such that the line connecting
the crosses follows the observational tendency.
}
\label{Fig-eq14}
\end{figure}

 To demonstrate that dilution by dust emission is not taking place, the
13.7\,$\mu$m profile is fitted with a Gaussian, and the Gaussian height
above the continuum is plotted as a function of colour
(Fig.~\ref{Fig-eq14}). The (pseudo-)continuum is estimated by linear
interpolation of the spectra between 11.5--12.0 and 14.6--15.1\,$\mu$m.  The
height from the Gaussian fit is divided by the value of the pseudo-continuum
at the Gaussian centre. Using this ratio minimises the influence of the
14.0~$\mu$m HCN, which is important because HCN is often found
among Galactic stars, but not in LMC stars.  The continuum should also avoid
the P- and R-branches of \CtHt\, as far as possible.  We measured both the
LMC spectra and the spectra of Galactic carbon stars reduced by
\citet{Hony02}.  The Galactic carbon stars are thought to be mostly located
in the Galactic disk. Stars with strong HCN or 13.9\,$\mu$m \CtHt, such as W
Ori, are not considered in this analysis, as the Gaussian fit failed.  With
the exception of IRAS 13416$-$6243, which is a post-AGB star exhibiting both
PAH bands and the 13.7\,$\mu$m \CtHt\, band, post-AGB stars and PNe are
also excluded from this discussion.

Neither the Galactic nor LMC carbon stars show a decrease of the 13.7\,$\mu$m 
\CtHt\, absorption band with redder infrared colour, i.e., with the optical
depth of the dust shell. Moreover, the largest absolute values of Gaussian
height tend to be found in stars with redder infrared colours among Galactic
stars.  The LMC stars show a smaller range in colour, but for the colour
range of overlap there is little evidence for any difference between LMC and
Galactic stars.  All of the LMC stars show \CtHt, but one Galactic star does
not (IRC+20\,326, Gaussian height of $-0.03$).

\section{ Discussion}


The shape of the 13.7~$\mu$m \CtHt\, feature is related to excitation
temperature as demonstrated by \citet{Aoki99} and \citet{Gautschy-Loidl04}. 
If the excitation temperature is about 1000\,K, there is a broad wing
longwards of 13.7\,$\mu$m (seen in S~Sct in Fig.\ref{Fig-spec14ISO}).  Among
our sample, there is no such a broad \CtHt\, band implying an excitation
temperature much lower than 1000~K.  Our observed spectra are reasonably
well reproduced by \CtHt\, at 500\,K (Fig.~\ref{Fig-spec14}).  This suggests
that these molecular bands are of circumstellar, rather than photospheric,
origin.

The fact that \CtHt\, becomes stronger rather than weaker towards redder
stars suggests that 13.7\,$\mu$m \CtHt\, originates outside of the region
where warm dust is emitting, and is therefore much less subject to infilling
by dust excess than are molecular bands from the photosphere.


\citet{Keady93} report that \CtHt\, can be detected throughout a shell, from
the photosphere up to the circumstellar envelope. Depending on the energy
level, the dominant molecules for the line formation are located in
different places in the photosphere or in the circumstellar envelope.  The
problem is that usually the inner hotter region shows a higher density and a
higher temperature, and therefore a higher optical depth. Thus, the inner
region will usually contribute more to the line formation.

The 13.7~$\mu$m \CtHt\,  feature thus provides two self-consistent arguments
for a molecular gas extended well into, and quite possibly beyond, the dust
forming region.  The gas is cool, and its features are not filled in by dust
emission, because the 13.7~$\mu$m band arises from a cool layer of gas above
the dust forming region.


The circumstellar \CtHt\, band strength does not show any particular
metallicity dependence. Despite the lower metallicity in the LMC, the \CtHt\
Gaussian fit depths are comparable to those found in Galactic stars. 
\citet{Sloan06} investigate carbon-rich stars in the Small Magellanic Cloud
(SMC), whose metallicity is probably about a quarter of the solar value.
They find that the \CtHt\, equivalent widths are higher among the SMC sample
than among their Galactic counterparts.  This may show the effect of carbon
synthesised inside the AGB star and brought the surface by the third
dredge-up.  Newly synthesised carbon is more important for the \CtHt\,
abundance than the effect of low initial metallicity and of low initial
carbon abundance.

The increase of the 13.7\,$\mu$m \CtHt\, depth with redder infrared colour
(Fig.~\ref{Fig-eq14}) indicates an increasing \CtHt\, column density.  This
is probably related to the mass in the circumstellar envelope.  There
are two possible reasons: the higher density may simply result in more
\CtHt\, or the higher density may actually facilitate the formation of \CtHt. 
There is a lot of scatter in a plot of 13.7\,$\mu$m \CtHt\, against colour,
suggesting that other effects such as abundance and temperature also
influence the band strength. Alternative explanation for the scatter could
include inner shocks caused by the pulsations \citep{Gautschy-Loidl04},
as found in some variability in the colour and \CtHt\, index \citep{Sloan06}.

There is no obvious detection of HCN among the LMC stars, apart from the
suspected feature at 14.0~$\mu$m in MSX LMC~601 and TRM~88.  While the
absence of HCN features could result from HCN emission filling in an
absorption feature, we consider this unlikely. \citet{Aoki99} and
\citet{Cernicharo99} detected HCN emission in some Galactic carbon stars
(only three stars out of the 9 stars they investigate), but one would
expect to see other HCN bands at other wavelengths. \citet{Matsuura05} and
\citet{vanLoon06} report few detections of the HCN band at 3.5~$\mu$m in LMC
stars compared to Galactic carbon stars.  Furthermore, only two stars in the
sample described here (NGC 978 IR4 and MSX LMC 494) show a suspected HCN
band at 7~$\mu$m.

The lower HCN abundance is most likely explained as a result of lower
abundance of nitrogen in a metal-poor environment.  Nitrogen is produced but
also destroyed in AGB stars and as a consequence will be much less abundant
with respect to carbon in a metal-poor environment.

The HCN 14.0~$\mu$m strength at a column density of
$3\times10^{17}$~cm$^{-2}$ is comparable to that of the 13.9~$\mu$m \CtHt\,
band at a column density of $1\times10^{18}$~cm$^{-2}$ at 500~K
(Fig.~\ref{Fig-spec14}). If the low HCN abundance is responsible for the
weak HCN features, the abundance of HCN could be less than one third of that
of \CtHt.

A chemical equilibrium model shows that the fraction of \CtHt\, in the
atmosphere exceeds that of HCN at C/O$\sim$1.3, and it will be three times
higher than that of HCN above C/O$>$1.7 \citep{Matsuura05}.  We assume that
the oxygen and nitrogen abundances are [O/H]\,=\,8.3, and [N/H]\,=\,7.5.
These elemental abundances are scaled from the solar elemental abundance
(log[n(C)/n(H)]+12=8.56 and log[n(N)/n(H)]=8.05 \citep{Anders89}) to the
metallicity of the LMC, for which we adopt a value [Fe/H]=$-0.6$.
The assumed oxygen and nitrogen abundances are comparable to those found in
red giants in the LMC \citep{Smith02}. The weak HCN but strong \CtHt\, bands
may show that LMC carbon stars have a carbon abundance of [C/H]\,=\,8.8--9.3.

A typical \CtHt\, column density of $1\times10^{18}$~cm$^{-2}$ can yield a
mass-loss rate of $dM/dt =3 \times 10^{-6} \times (r_{{\rm in}}/R_{*})$
M$_{\odot}$\,yr$^{-1}$. The parameter $r_{{\rm in}}$/$R_{*}$ is the inner
radius of the \CtHt\, line-forming region with respect to the stellar radius. 
We assume that the \CtHt\,
abundance is 1$\times10^{-5}$ with respect to H$_2$ \citep{Matsuura05}, that
the expansion velocity is 20~km\,s$^{-1}$, and that the stellar radius
$R_{*}$ is 300 R$_{\odot}$.  We also assume that the outer radius of the
\CtHt\, circumstellar shell is much larger than the inner radius, and that
the outer radius can be ignored for this calculation.  If most of the
\CtHt\, is formed at the photosphere ($r_{{\rm in}}/R_{*}$=1), the mass-loss
rate would be $dM/dt=3\times 10^{-6} M_{\odot}$\,yr$^{-1}$. This would be
the minimum gas mass-loss rate of our sample.  However, the line-forming
region is much further out, $r_{{\rm in}}$/$R_{*}$ would be larger than 1.
500~K corresponds to about 16 stellar radii or more \citep{Hoefner03}.  This
would be equivalent to a gas mass-loss rate of 5$\times 
10^{-5}$~M$_{\odot}$\,yr$^{-1}$ or higher.

van Loon et al. (2006) have estimated the mass-loss rate of 14 stars out of
26 in our sample by modelling the spectral energy distribution (SED). The
mass-loss rate range from 9$\times 10^{-6}$~M$_{\odot}$\,yr$^{-1}$
(NGC~1978-IR1) to 5$\times 10^{-5}$~M$_{\odot}$\,yr$^{-1}$ (MSX~LMC~635).
Thus the two independent mass-loss rates, from our \CtHt\, and from the SED,
are consistent.
\citet{vanLoon06} use a gas-to-dust ratio of 500, based
on the lower metallicity.  These consistent mass-loss rates estimated from
the \CtHt\, band and the SED may be supportive of such a high gas-to-dust ratio
at low metallicity \citep{vanLoon00}.




 


\section{Conclusion}

 We analyse molecular bands found in Spitzer spectra of LMC carbon stars.
\CtHt\, bands are detected at 7.5 and 13.7~$\mu$m.
The 13.7~$\mu$m \CtHt\, bands are of circumstellar origin. This is 
demonstrated by the absence of infilling of this band by dust emission, even
amongst stars with thick circumstellar envelopes, and because the excitation
temperatures of \CtHt\, appear to be about 500~K. There are abundant \CtHt\,
molecules despite the lower metallicity in the LMC compared to our Galaxy.
This could be explained by carbon enrichment in the AGB stars. In contrast
to \CtHt\, there is no clear evidence for HCN bands among the LMC sample,
implying that low nitrogen abundances affect the HCN abundance.

We estimate from \CtHt\, gas mass-loss rates ranging from 3$\times
10^{-6}$~M$_{\odot}$\,yr$^{-1}$ to 5$\times 10^{-5}$~M$_{\odot}$\,yr$^{-1}$.  This
is consistent with mass-loss rate estimated from the SED.


\section{acknowledgements}
We appreciate informative comments from the referee Dr. U.G. J{\o}rgensen.
A.A.Z., M.M. and E.L. are financially supported by  PPARC. 
M.M. thanks the IRS Team at Cornell University for their hospitality during her stay there.
This visit was supported by the
Peter Allen Travelling Grant of University of Manchester.
M.M. is JSPS Research Fellow.
Support for G.C.S. was provided by NASA through Contract
Number 1257184 issued by the Jet Propulsion Laboratory,
California Institute of Technology under NASA contract 1407.
P.R.W. has been partially supported by a Discovery Grant 
from the Australian Research Council

\label{lastpage}


\begin{thebibliography}{}

\bibitem[Anders \& Grevesse(1989)]{Anders89}  
  Anders E., Grevesse N., 1989, Geochimica et Cosmochimica Acta 53, 197

\bibitem[Aoki et al.(1998)]{Aoki98}
  Aoki W., Tsuji T., Ohnaka K., 1998, A\&A 340, 222

\bibitem[Aoki et al.(1999)]{Aoki99}
  Aoki W., Tsuji T., Ohnaka K.,
  1999, A\&A 350, 945

\bibitem[Carter et al.(1993)Carter, Mills \& Handy]{Carter93}
  Carter S., Mills I.M., Handy N.C. 1993,
  Journal of Chemical Physics, 99, 4379

\bibitem[Cernicharo et al.(1999)]{Cernicharo99}
  Cernicharo J., Yamamura I., Gonz\'{a}lez-Alfonso E., de Jong T.,
  Heras A., Escribano R., Ortigoso J.,
  1999, ApJ, 526, L41

\bibitem[Durai and Hatchell(2000)]{Duari00}
   Duari D., Hatchell J., 2000, A\&A, 358, L25

\bibitem[Gautschy-Loidl et al.(2004)]{Gautschy-Loidl04}
  Gautschy-Loidl R., H\"{o}fner S., J{\o}rgensen U.G., Hron J.,
  2004, A\&A 422, 289

\bibitem[H\"{o}fner et al.(2003)]{Hoefner03}
 H\"{o}fner S., Gautschy-Loidl R., Aringer B., J{\o}rgensen U.G.,
  2003, A\&A 399, 589

\bibitem[Hony et al.(2002)]{Hony02}
  Hony S., Waters L.B.F.M., Tielens A.G.G.M.,
  2002, A\&A 390, 533

\bibitem[Houck et al.(2004)]{Houck04}
  Houck J.R., Roellig T.L., van Cleve J., et al.,
  2004, ApJS 154, 18

\bibitem[Hron et al.(1998)]{Hron98}
  Hron J., Loidl R., H\"{o}fner S., J{\o}rgensen U.G., Aringer B., Kerschbaum F.,
  1998, A\&A 335, L69

\bibitem[Jacquemart et al.(2003)]{Jacquemart03}
  Jacquemart D., Mandin J.-Y., Dana V., et al., 2003, 
  Journal of Quantitative Spectroscopy \& Radiative Transfer 82, 363

\bibitem[Keady \& Hinkle (1988)]{Keady88}
  Keady J.J., Hinkle K.H.
  1988, ApJ 331, 539

\bibitem[Keady \& Ridgway (1993)]{Keady93}
  Keady J.J., Ridgway S.T.,
  1993, ApJ 406, 199

\bibitem[Kraemer et al.(2002)]{Kraemer02}
  Kraemer K.E., Sloan G.C., Price S.D., Walker H.J.,
  2002, ApJS 140, 389

\bibitem[Matsuura et al.(2002a)]{Matsuura02a}
  Matsuura M., Yamamura I., Cami J., Onaka T., \& Murakami H.,
  2002a, A\&A 383, 972

\bibitem[Matsuura et al.(2002b)]{Matsuura02b}
  Matsuura M., Zijlstra A.A., van Loon J.Th., et al.,
  2002b, ApJ 580, L133

\bibitem[Matsuura et al.(2005)]{Matsuura05}
  Matsuura M., Zijlstra A.A., van Loon J.Th., et al.,
  2005, A\&A 434, 691

\bibitem[Millar et al.(2000)]{Millar00}
  Millar T.J., Herbst E., Bettens R.P.A.,
  2000, MNRAS 316, 195

\bibitem[Rothman et al.(2005)]{Rothman05}
  Rothman L.S., Jacquemart D., Barbe A., et al., 2005,
  Journal of Quantitative Spectroscopy \& Radiative Transfer 
   96, 139

\bibitem[Sloan et al.(2006)]{Sloan06}
  Sloan G.C., Kraemer K.E., Matsuura M., Wood P.R.,
  Price S.D., Egan M.P., 2006, ApJ 645, in press

\bibitem[Smith et al.(2002)]{Smith02}
  Smith V.V., Hinkle K.H.. Cunha K. et al.,
  2002, AJ 124, 3241

\bibitem[Smith \& Lambert(1985)]{Smith85}
  Smith V.V., Lambert D.ÊL.,
  1985, ApJ 294, 326

\bibitem[Tsuji (1981)]{Tsuji81}
  Tsuji T., 1981, JApA 2, 253


\bibitem[van Loon et al.(2000)]{vanLoon00}
  van Loon J.Th., 2000, A\&A 354, 125

\bibitem[van Loon et al.(1999)]{vanLoon99b}
  van Loon J.Th., Zijlstra A.A., Groenewegen M.A.T.,
  1999, A\&A 346, 805

\bibitem[van Loon et al.(2006)]{vanLoon06}
   van Loon J.Th.,  Marshall J.R., Cohen M., 
   Matsuura M., Wood P.R., Yamamura I., Zijlstra A.A.,
   2006, A\&A 447, 971

\bibitem[Zijlstra et al.(2006)]{Zijlstra06}
  Zijlstra A.A., Matsuura M., van Loon J.Th., et al., 2006,
  MNRAS, accepted


\end{thebibliography}
\end{document}